\documentclass[prb,showpacs,floatfix,twocolumn,letterpaper]{revtex4}

\usepackage{graphicx}

\newcommand{\cco}{CuCr$_2$O$_4$}
\newcommand{\nco}{NiCr$_2$O$_4$}
\begin{document}

\title{Spin-induced symmetry breaking in orbitally ordered \nco\/ and \cco}

\author{Matthew R. Suchomel}\email{suchomel@aps.anl.gov}
\author{Daniel P. Shoemaker}\email{dshoemaker@anl.gov}
\author{Lynn Ribaud}\email{ribaud@anl.gov}
\affiliation{X-Ray Science Division and Material Science Division\\ 
Argonne National Laboratory, Argonne IL, 60439, USA}
\author{Moureen C. Kemei}\email{kemei@mrl.ucsb.edu}
\author{Ram Seshadri}\email{seshadri@mrl.ucsb.edu}
\affiliation{Materials Department and Materials Research Laboratory\\
University of California, Santa Barbara, CA, 93106, USA}

\date{\today}

\begin{abstract}
At room temperature, the normal oxide spinels \nco\/ and \cco\/ are 
tetragonally distorted and crystallize in the $I4_1/amd$ space group
due to cooperative Jahn-Teller ordering driven by the orbital degeneracy
of tetrahedral Ni$^{2+}$ ($t_2^4$) and Cu$^{2+}$ ($t_2^5$).
Upon cooling, these compounds undergo magnetic ordering transitions; interactions being
somewhat frustrated for \nco\/ but not for \cco.
We employ variable-temperature high-resolution synchrotron X-ray powder 
diffraction to establish that at the magnetic ordering temperatures
there are further structural changes, which result in both compounds  
distorting to an orthorhombic structure consistent with the $Fddd$ space
group. \nco\/ exhibits additional distortion, likely within the same space 
group, at a yet-lower transition temperature of $T$ = 30\,K. The tetragonal to 
orthorhombic structural transition in these compounds appears to primarily 
involve changes in NiO$_4$ and CuO$_4$ tetrahedra. 
\end{abstract}

\maketitle 

\section{Introduction} 
Strong coupling between spin, lattice, and orbital degrees of freedom in
functional transition metal oxide compounds results in rich behavior such as
the tendency for cooperative Jahn-Teller\cite{dunitz_1957} and spin-Peierls
distortions.\cite{beni_1972} Such coupling between the different degrees of
freedom enables multifunctionality as observed in multiferroics
$R$MnO$_3$($R$ = late rare earth).\cite{Fabreges_2009,lee_2008}  
In these systems, manipulation of one property can influence another,
exemplified by the electric field control of magnetic polarization in
HoMnO$_3$\cite{Lottermoser_2004}. Seeking out such strong links between
distinct degrees of freedom represents a powerful strategy in the search for
new  multifunctional systems, and affords unique opportunities for a deeper
understanding of these interactions.\cite{Glazkov_2009, Ueda_2005} 

One such frequently studied interaction is magnetostructural coupling in
geometrically frustrated antiferromagnets
\cite{tchernyshyov_2004,rudolf_2007} where a structural distortion lifts the
large ground state degeneracy allowing long range magnetic
order.\cite{kant_2010,lee_2007} However, frustration-driven magnetostructural 
coupling is not expected in the ferrimagnetic spinels with the formula
$A$Cr$_2$O$_4 $ where $A$ is a magnetic cation. This is a consequence of the
magnetic $A$-O-Cr$^{3+}$ interaction usually being collectively  
stronger than the frustrated interactions between the Cr$^{3+}$.
Furthermore, Jahn-Teller activity of the $A$ site cation can cause tetragonal 
distortions that should further alleviate frustration in the Cr$^{3+}$ 
sublattice. Nonetheless, previous structural, thermodynamic, and magnetic 
studies of \nco\cite{ishibashi_2007,Klemme_2002} report a coupled magnetic 
and structural transition, and infrared spectroscopy measurements suggest
concurrent magnetic and structural transitions  in
\cco.\cite{bordacs_2009} 

Structural transitions at the magnetic ordering temperatures have been 
observed in numerous transition metal oxide antiferromagnets such as 
Cr$_2$O$_3$,\cite{Greenwald1951} MnO,\cite{smart_1951,Roth1958} 
FeO,\cite{smart_1951,Roth1958} CoO,\cite{smart_1951,Roth1958} and 
NiO.\cite{smart_1951,Roth1958} Cubic to rhombohedral transformations 
are found in MnO, FeO, and NiO, while CoO
undergoes a cubic to tetragonal transition. The rhombohedral lattice
constants of Cr$_2$O$_3$ change at its antiferromagnetic ordering
temperature. Two mechanisms of magnetostructural coupling have been suggested 
in these compounds based on neutron and X-ray diffraction measurements. 
Li has suggested that magnetostructural coupling in NiO, MnO, CoO, and FeO
is driven by magnetostriction,\cite{Li1955} where anisotropy arises from the
selection of a magnetic ordering axis and drives the magnetocrystalline
deformation. Smart and Greenwald alternatively proposed that distortions in
the above binary oxides are caused by exchange striction, which is the 
displacement of interacting ions to strengthen exchange coupling thus 
modifying the underlying lattice.\cite{smart1950} The relations between crystal
distortions and exchange interactions are challenging to identify. For
example, it is difficult to find a unique solution to certain magnetic
scattering patterns.\cite{Li1955,Roth1958}

Here, we determine the low temperature structures of \nco\/ and \cco\/ across
the transitions associated with magnetic ordering using high resolution
synchrotron powder X-ray diffraction. These compounds are fully-ordered and
stoichiometric normal cubic spinels with the space group $Fd\bar 3m$ at 
temperatures above 320\,K\cite{crottaz_1997,ohgushi_2008} for \nco\/ 
and 853\,K\cite{Ye_1994,ohgushi_2008} for \cco.\cite{chukalkin_1985} 
Cr$^{3+}$  3$d^3$ preferentially populates the octahedral sites 
because of the strong crystal field stabilization of the half occupied
nondegenerate $t_{2g}$ states and empty $e_g$ states, while Ni$^{2+}$ 3$d^8$
and Cu$^{2+}$ 3$d^9$ are found on the tetrahedral sites.\cite{dunitz_1957} The
tetrahedral crystal field around Ni$^{2+}$ 3$d^8$ and Cu$^{2+}$ 3$d^9$ in the
cubic phase results in fully occupied low energy $e$ levels and triply
degenerate $t_2$ levels rendering this structure potentially
unstable.\cite{kanamori_1960,gerloch_1981} A cooperative lattice distortion --
from cubic to tetragonal symmetry -- lifts the orbital degeneracy in \nco\/ at
320\,K\cite{Klemme_2002,crottaz_1997,tovar_2006} and in \cco\/ at
853\,K.\cite{Ye_1994,tovar_2006} There had been a debate in the literature
concerning the ambient temperature structure of \cco. Using neutron
diffraction data, Prince postulated that the noncentrosymmetric space group
$I\bar42d$ was a better structural fit than
$I4_1/amd$.\cite{prince_1957} More recently, Dollase and O'Neill showed no
statistically significant advantage to using the $I\bar42d$ structural
model over the centrosymmetric structure $I4_1/amd$.\cite{dollase_1997} In the tetragonal structure of \cco\/, CuO$_4$ tetrahedra are compressed toward a square planar configuration thus lifting orbital degeneracy. \cite{dunitz_1957} The
tetragonal structure of \nco\/ is known to crystallize in the space group
$I4_1/amd$ with elongated NiO$_4$ tetrahedra.\cite{Ueno_1999} Previous work has also shown further distortion of
tetragonal \nco\/ to an orthorhombic phase, which occurs at the magnetic
transition temperature $T_N$ = 60\,K and has been observed in thermodynamic,
X-ray diffraction, and magnetic studies.\cite{Klemme_2002,ishibashi_2007}

Noncollinear ferrimagnetism that is not described by the N\'{e}el model is
observed in both \nco\, and \cco. Tomiyasu and Kagomiya describe a magnetic
structure comprising of a ferrimagnetic component and an antiferromagnetic
component in \nco.\cite{tomiyasu_2004,Klemme_2002} These authors used neutron
scattering to show that the antiferromagnetic component orders
at $T$ = 31\,K while the ferrimagnetic component orders at $T$ = 74\,K. A
saturation magnetization moment of 0.3\,$\mu_B$ \textit{per} formula unit has 
been reported for \nco.\cite{tomiyasu_2004} Neutron scattering studies 
on \cco\/ suggest a magnetic structure comprising of two canted Cr$^{3+}$ 
sublattices with a net moment, and Cu$^{2+}$ sublattice that couples 
antiferromagnetically to the net moment of the Cr$^{3+}$ sublattices below 
$T_N$ = 135\,K.\cite{prince_1957,tovar_2006,ohgushi_2008} The magnetic
moment of \cco\/ in this structure is 0.5\,$\mu_B$ \textit{per} formula unit.

Given this prior evidence of concurrent magnetic and structural transitions in
\nco\cite{ishibashi_2007,Klemme_2002} and \cco\cite{bordacs_2009}, there
is a clear need for further exploration of these compounds.  In this 
study, we employ high-resolution temperature-dependent powder X-ray
diffraction, magnetic susceptibility, and heat capacity measurements to
investigate magnetostructural coupling in \nco\/ and \cco. This is the first
observation by X-ray powder diffraction of the tetragonal to orthorhombic
structural distortion of \cco\/ at the ferrimagnetic ordering temperature. We
also reveal for the first time X-ray diffraction evidence of further symmetry
lowering in orthorhombic \nco\/ at the second magnetic transition
$T$ = 30\,K. These results affirm strong magnetostructural coupling
can also occur in spinels that are not expected to be frustrated. 
This new understanding of coupling between spin and lattice degrees of freedom 
in \nco\/ and \cco\/ suggests that these compounds are promising 
magnetodielectrics, and provides considerable motivation for further 
investigation of magnetostructural coupling in related spinel compounds.

\section{Methods}

\nco\/ was prepared by dissolving stoichiometric amounts of
Ni(NO$_3$)$_2\cdot$6H$_2$O and Cr(NO$_3$)$_3\cdot$9H$_2$O in deionized water.
The nitrate solution was heated to evaporate the solvent, leaving a
precipitate that was ground and calcined at 1000$^{\circ}$\,C for 24 hours. A
dark green powder of \nco\/ was obtained. Black shiny single crystals of
\cco\/ were prepared following the flux method described by 
Ye \textit{et al.}\cite{Ye_1994} 
K$_2$Cr$_2$O$_7$ was used as a reactive flux that partly decomposes to 
Cr$_2$O$_3$ at $\sim$ 700\,K.  Ye \textit{et al.} propose that the
reduction of Cr$^{6+}$ into Cr$^{3+}$ plays an important role in stabilizing
the oxidation state of Cu$^{2+}$ during the synthesis of CuCr$_2$O$_4$.
\cite{Ye_1994}  K$_2$Cr$_2$O$_7$ acts both as a flux and a source of
Cr$_2$O$_3$. A 20\,g mixture of 17.8\% mass CuO (Sigma Aldrich 98$\%$) and
82.2\% mass K$_2$Cr$_2$O$_7$ (Fisher 99$\%$) with 0.2\,g Bi$_2$O$_3$ as a 
second flux was prepared. The mixture was ground using an agate mortar and 
pestle, placed in a covered platinum crucible, heated to 800$^{\circ}$\,C with 
a ramp of 100$^{\circ}$C/h, held for 24\,h, and slowly cooled to ambient 
temperature at 15$^{\circ}$C/h. After the reaction, black crystals of 
CuCr$_2$O$_4$ were collected and washed in boiling water. It should be noted 
that more conventional solid state preparation yielded samples with 
significantly broader linewidths in the synchrotron X-ray diffraction profile, 
potentially obscuring the ability to fully characterize the low-temperature 
structure.

High-resolution ($\Delta d/d \approx$ $10^{-4}$) synchrotron X-ray powder
diffraction data were recorded on beamline 11-BM at the Advanced Photon
Source (APS), Argonne National Laboratory.\cite{wang_2008} Scans were
collected using a 2$\Theta$ step size of 0.001$^{\circ}$ with
$\lambda$ = 0.413441\,\AA\/ for \nco\, and $\lambda$ = 0.41326\,\AA\/ for 
\cco\/ in a closed-flow helium cryostat over the temperature range 7\,K to
300\,K. The sample was in direct contact with the helium 
exchange gas during data collection, and was spun at 5\,Hz to improve powder 
averaging. Structural models of \nco\/ and \cco\/ were fit to the diffraction 
data using the Rietveld refinement method as implemented in the EXPGUI/GSAS 
software program.\cite{toby_expgui_2001, larson_2000} Crystal structures were
visualized using the program VESTA.\cite{momma_vesta_2008} Both samples
reported here had a small, second impurity phase that was also quantitatively 
fit using the Rietveld method. The \nco\/ sample was determined to have 
a 0.5\,wt.-\% of Cr$_2$O$_3$, and the \cco\/ sample a  1.1\,wt.-\% CuO
impurity.

Magnetic susceptibility measurements on powder samples were performed using a 
Quantum Design MPMS 5XL superconducting quantum interference device (SQUID)
magnetometer. Heat capacity measurements were collected on pellets of 50$\%$
mass silver and 50$\%$ mass sample using a Quantum Design Physical Properties
Measurement System. The pellets were prepared by grinding equal amounts of
silver and sample in an agate mortar and pestle followed by pressing at
$\sim$ 330\,MPa. Apiezon N grease was used to enhance thermal coupling
between the sample and the stage. The heat capacity of the Apiezon N grease and
silver were collected separately and subtracted from the measured heat capacity.

\section{Results and Discussion}

\subsection{Magnetism}

\begin{figure}
\centering\includegraphics[width=9cm]{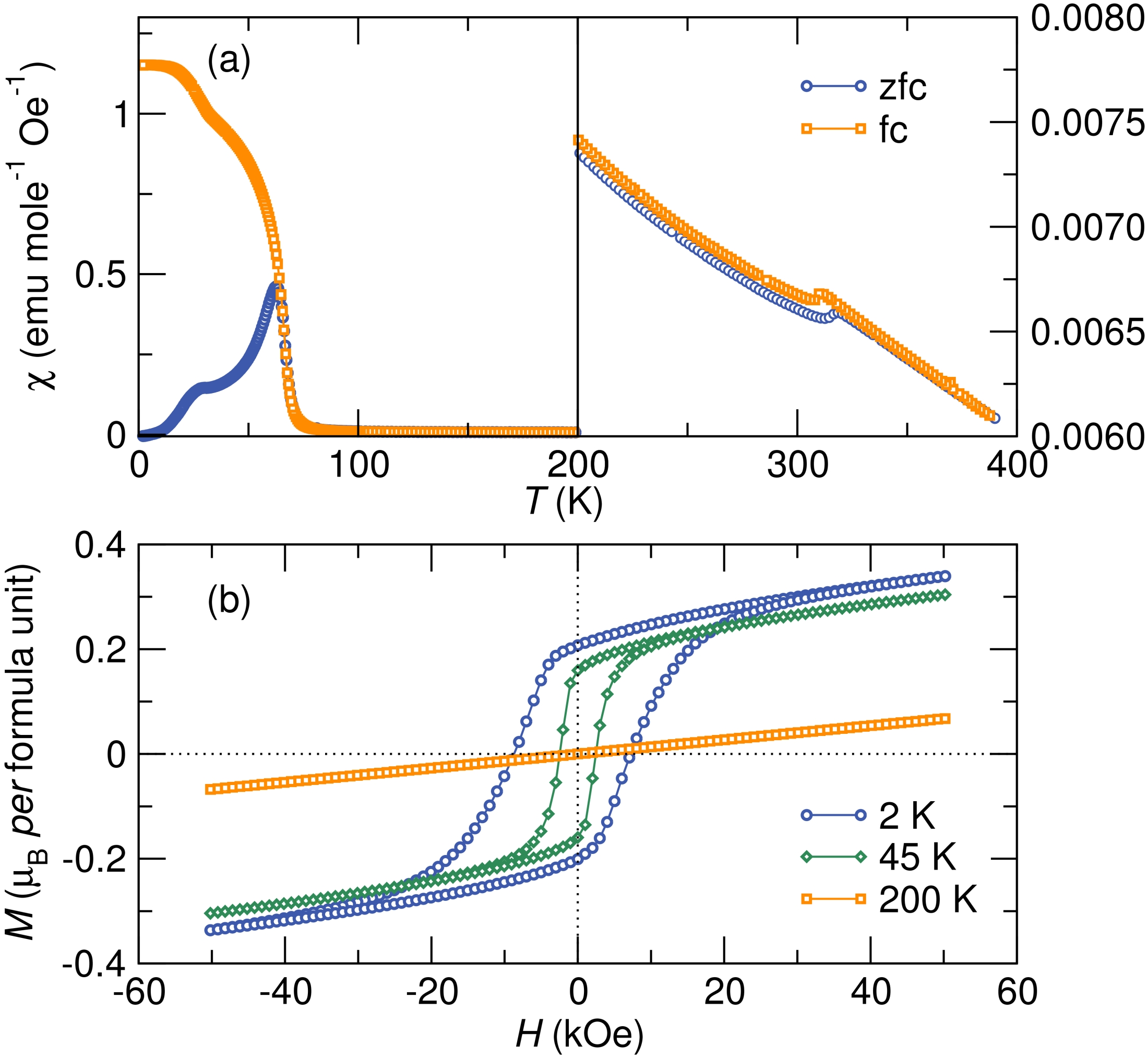}\\
\caption{(Color online) Magnetic measurements of the spinel \nco. (a) Zero
field cooled and field cooled temperature
dependent magnetic susceptibility measured under a 1000\,Oe DC field show three
anomalies at 310\,K, 65\,K and 30\,K. \nco\/ displays little change in the 
magnetism at 310\,K, and is seen to  order ferrimagnetically at
65\,K, with an additional change in the magnetic structure at 30\,K.
(b) The isothermal field dependent magnetization measured above the magnetic
ordering temperature shows paramagnetic behavior. At 2\,K, the coercive field 
and saturation magnetization are significantly larger than what is observed 
at 45\,K.} 
\label{fig:magnco}
\end{figure}

\begin{figure}
\centering\includegraphics[width=8cm]{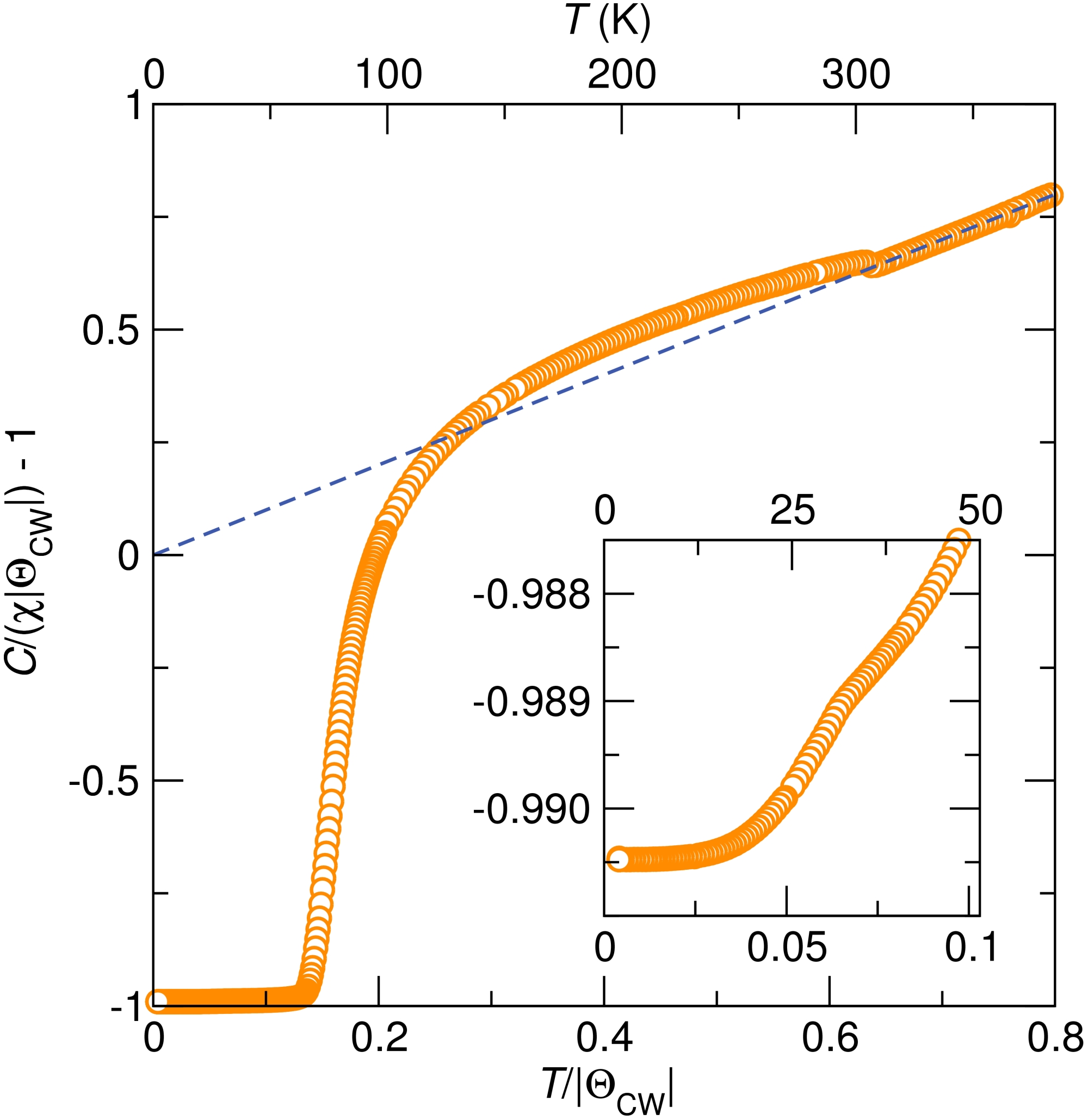}\\
\caption{(Color online) Normalized inverse magnetic susceptibility of \nco\/
showing ideal Curie Weiss paramagnetism above 310\,K. Weak compensated
interactions arise at 310\,K and persist to about 65\,K below which strong
uncompensated interactions dominate. The subtle magnetic transition at 30\,K
is shown in the inset.} 
\label{fig:cwnco}
\end{figure}

Three magnetic transitions are observed in the temperature dependent magnetic
susceptibility of \nco\/ (Fig.\,\ref{fig:magnco}). A high temperature
transition occurs at 310\,K where cooperative Jahn-Teller distortions lift
the orbital degeneracy in \nco\/ and lower the structural symmetry from cubic
($Fd\bar{3}m$) to tetragonal ($I4_1/amd$) [Fig.\,\ref{fig:magnco} (a)].
Weak, compensated magnetic interactions occur at 310\,K, as illustrated by
the scaled inverse susceptibility of \nco\/ (Fig.\,\ref{fig:cwnco}). 
The scaling is carried out by recasting the Curie-Weiss equation 
using:\cite{melot_CW}

\begin{equation}
\frac{C}{\chi|\Theta_{CW}|} + \mbox{sgn}(\Theta_{CW}) = \frac{T}{|\Theta_{CW}|}
\end{equation}

\noindent 
The linear dependence of the magnetization on the applied field at 200\,K 
[Fig.\,\ref{fig:magnco}(b)] suggests that \nco\/ is mainly paramagnetic down to
65\,K where there is a transition to a ferrimagnetic state 
[Fig.\,\ref{fig:magnco}(a)]. 
The normalized inverse magnetic susceptibility trace 
shows the development of strong uncompensated magnetic correlations at 65\,K
(Fig.\,\ref{fig:cwnco}). A small coercive field and saturation magnetization
is observed in the field dependent magnetization of \nco\/ at 45\,K 
[Fig.\,\ref{fig:magnco}(b)] in agreement with the onset of ferrimagnetic order.
Tomiyasu and Kagomiya attribute the magnetic transition at 65\,K in \nco\/ to
the ordering of the longitudinal ferrimagnetic component of
\nco.\cite{tomiyasu_2004} At 30\,K, another anomaly is observed in both zero field cooled (ZFC)
and field cooled (FC) measurements of the temperature dependent magnetic susceptibility
[Fig.\,\ref{fig:magnco}(a)] as well as in the scaled inverse susceptibility
(Fig.\,\ref{fig:cwnco}) of \nco. Below 30\,K, an increase in the coercive
field and the saturation magnetization of \nco\/ is observed 
[Fig.\,\ref{fig:magnco}(b)]. Previous neutron diffraction measurements of 
\nco\/ attribute this anomaly to the ordering of the antiferromagnetic 
component of \nco.\cite{tomiyasu_2004}

\begin{figure}
\centering\includegraphics[width=8cm]{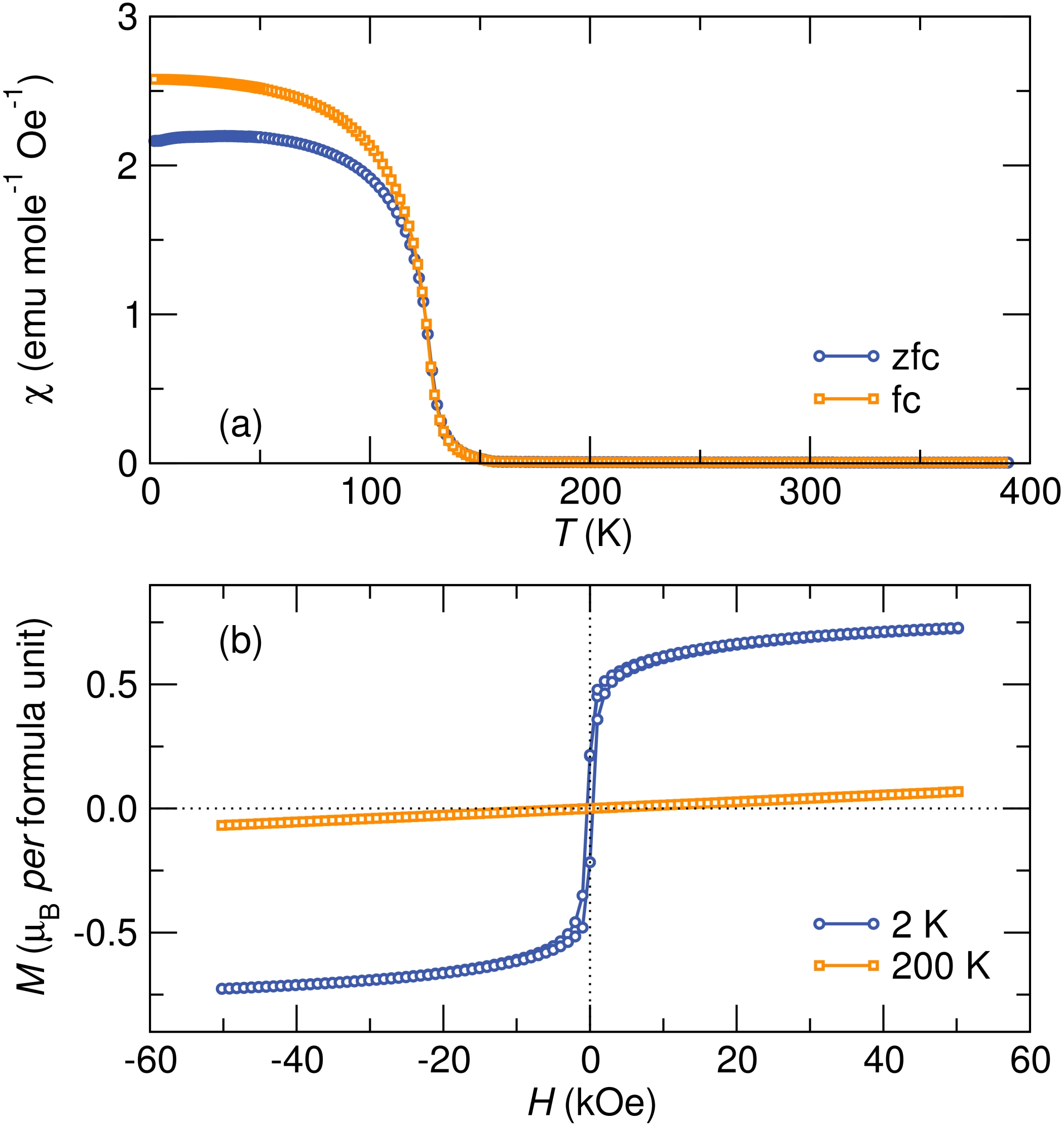}\\
\caption{(Color online) Magnetic measurements of the spinel \cco. (a)
Magnetic susceptibility as a function of temperature under a 1000\,Oe DC field
shows an increase in susceptibility at the magnetic ordering temperature
$\approx$130\,K in both zero field cooled and field cooled
measurements. This is a paramagnetic to ferrimagnetic
transition.  (b) Isothermal field dependent magnetization measured above
(200\,K) and below (2\,K) the magnetic ordering temperature.} 
\label{fig:magcco}
\end{figure}

The temperature dependent magnetic susceptibility of \cco\/ shows a rapid
increase at 130\,K where there is a paramagnetic to ferrimagnetic transition
[Fig.\,\ref{fig:magcco}(a)]. The ZFC susceptibility
exhibits a reduced low temperature saturation value when compared to the
FC susceptibility data illustrating domain behavior. 
A linear dependence of magnetization with applied field
occurs before the onset of magnetic order while a magnetization trace with a
coercive field of 380\,Oe and a saturation magnetization of 0.725\,$\mu_B$ is
measured at 2\,K. The measured saturation magnetization of \cco\/ is in good
agreement with that of the triangular magnetic structure observed previously using
neutron powder diffraction.\cite{prince_1957}

The Curie-Weiss (CW) equation $\chi = C/(T-\Theta_{CW})$ is applied to
paramagnetic regimes of \nco\/ and \cco\/ yielding an effective moment
($\mu_{eff}$) of 6.53 $\mu_B$ \textit{per} formula unit for \nco\/ and
4.27\,$\mu_B$ \textit{per} formula unit for \cco. The expected 
$\mu_{eff}$ of \nco\/ is 6.16\,$\mu_B$ \textit{per} formula unit of 
\nco. This value is slightly smaller that the experimentally determined value of 
6.53\,$\mu_B$ \textit{per} formula unit obtained from fitting the paramagnetic 
regime to the Curie-Weiss model, implying a small orbital contribution to the 
measured moment. The expected $\mu_{eff}$ of 5.74\,$\mu_B$ \textit{per} formula unit 
of \cco\/ is much larger than the experimental value suggesting the likely 
presence of magnetic correlations in the paramagnetic regime.\cite{kemei2012} 
The Weiss temperature ($\Theta_{CW}$) of \nco\/ is $-$487\,K while that of 
\cco\/ is $-$147\,K. The frustration index ($|\Theta_{CW}|/T_N$) of \nco\/ 
is about 7.8 and that of \cco\, is 1.1 indicating that \nco\/ is the more 
frustrated compound. The negative sign of $\Theta_{CW}$ coupled with the low 
saturation magnetization observed in isothermal field dependent measurements 
is consistent with noncollinear ferrimagnetic ordering in \nco\/ and \cco.

\begin{figure}
\centering\includegraphics[width=8cm]{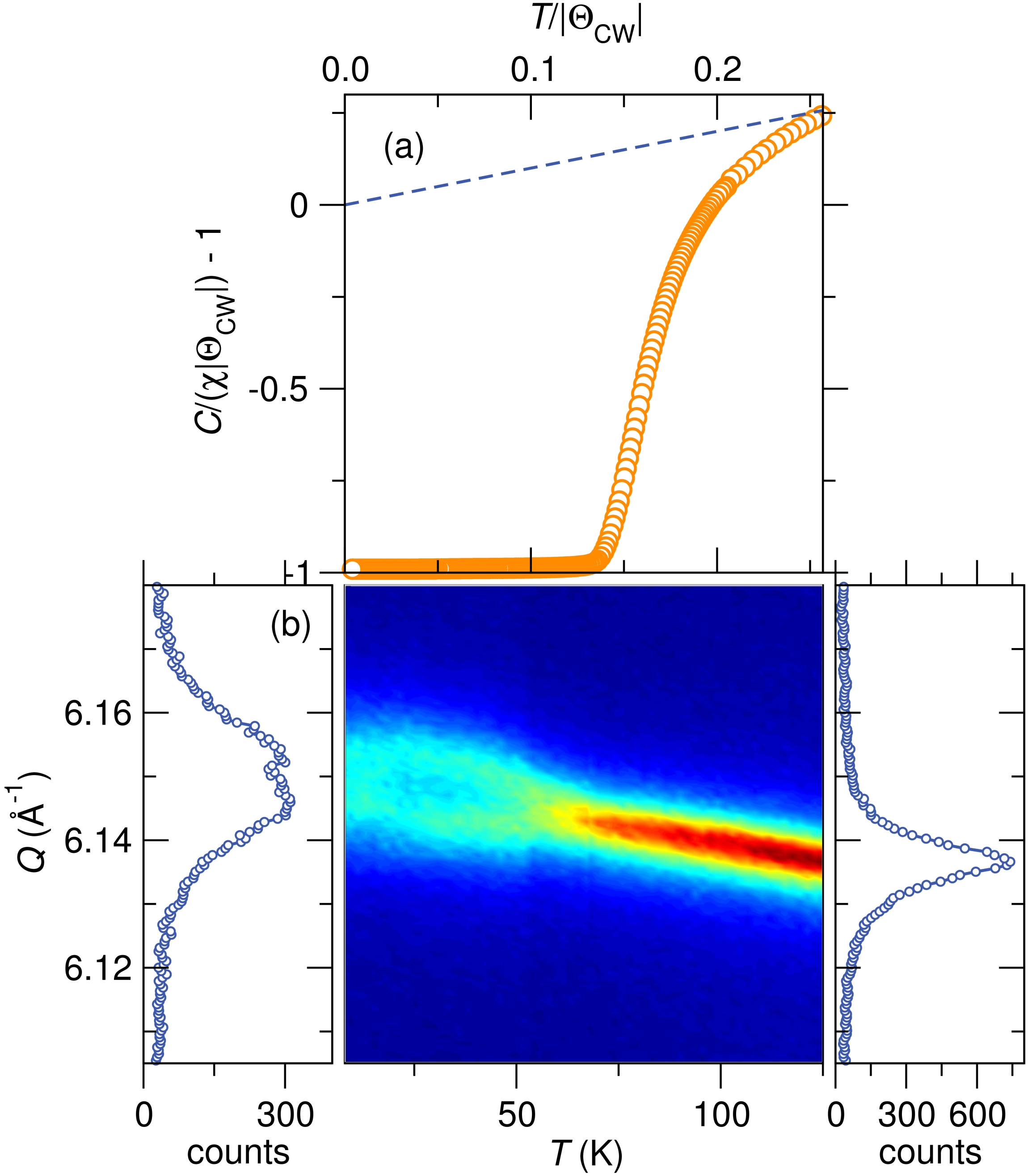}\\
\caption{(Color online) Magnetostructural coupling in \nco. (a) \nco\/ orders
ferrimagnetically at $T_N$\,=\,65\,K, where the normalized inverse magnetic
susceptibility deviates negatively from ideal Curie-Weiss
paramagnetic behavior. (b) A structural transition occurs
at the ferrimagnetic ordering temperature seen from the
splitting of the tetragonal 440 diffraction peak into 080 and 800
orthorhombic peaks. Below 30\,K, a subtle peak narrowing and intensity change
is coincident with anomalies in magnetic and specific heat measurements.
125\,K and 7\,K diffraction patterns are shown to the right and left of the
central panel.} 
\label{fig:vtnco}
\end{figure}

\begin{figure}
\centering\includegraphics[width=8cm]{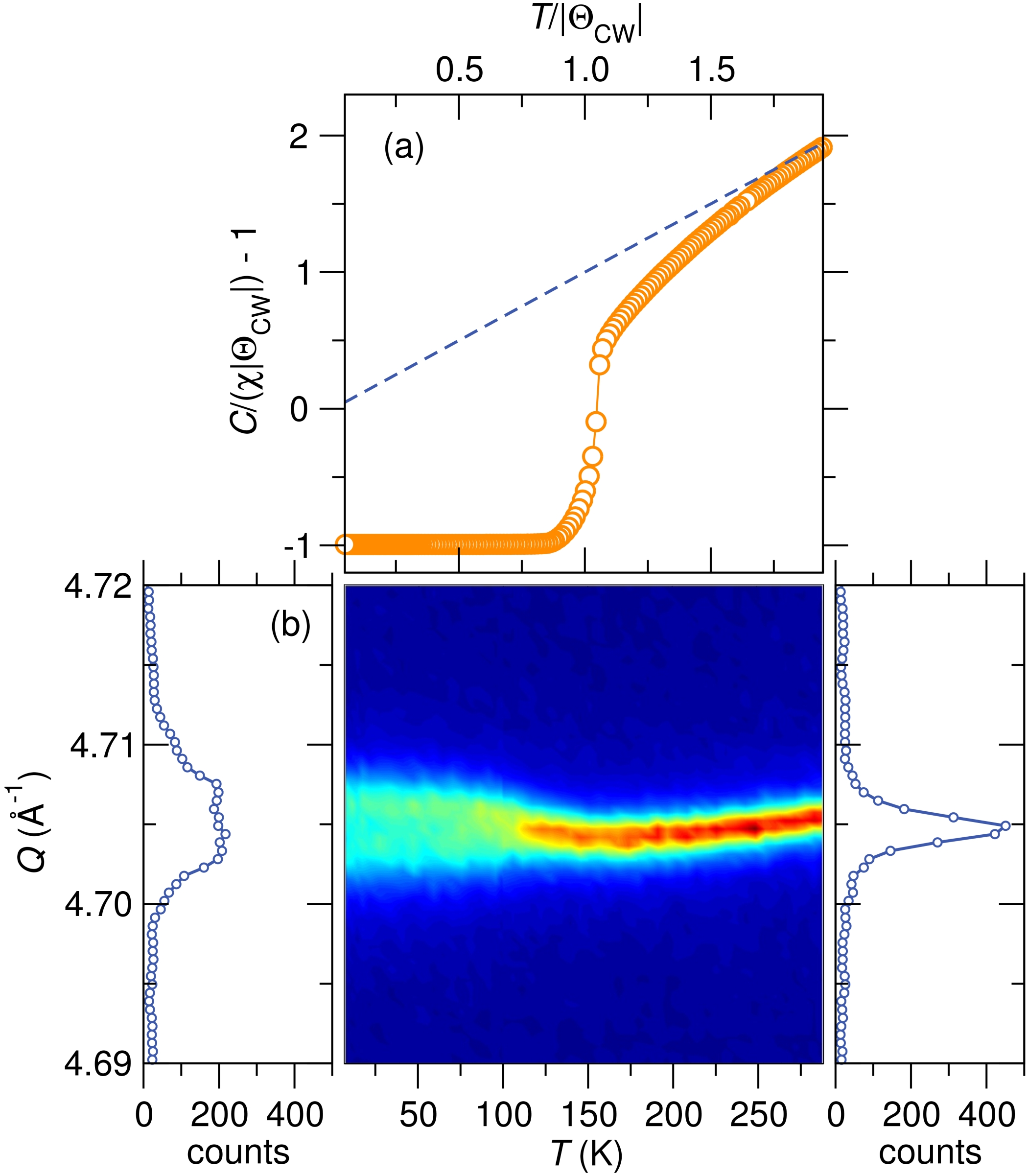}\\
\caption{(Color online) Magnetostructural coupling in \cco. (a) Long-range
ferrimagnetic order occurs at $T_N$ = 130\,K in \cco\/ where the normalized
inverse magnetic susceptibility of \cco\/ deviates negatively from
ideal Curie-Weiss behavior (b) Concurrent with the onset of
magnetic order is a structural transition seen in the 
splitting of the tetragonal 322 reflection into orthorhombic 206
and 260 reflections. Diffraction patterns at 288\,K and 7\,K are shown
to the right and left of the central plot respectively.} 
\label{fig:vtcco}
\end{figure}

The magnetic transitions of \cco\/ and \nco\/ are strongly coupled to the
lattice. All magnetic changes in \nco\/ are accompanied by structural
transitions. The known Jahn-Teller cubic to tetragonal structural distortion
in \nco\/ at 310\,K causes a small change in the temperature dependent
magnetization [Fig.\,\ref{fig:magnco}(a)].\cite{ishibashi_2007} Ishibashi
and Yasumi reported further distortion from tetragonal to orthorhombic
symmetry at the onset of ferrimagnetic order 
($T_N$ = 65\,K).\cite{ishibashi_2007} 
We observe this tetragonal to orthorhombic crystal
distortion occurring concurrently with the onset of ferrimagnetic order in
\nco\/ in Fig.\,\ref{fig:vtnco}. A low temperature anomaly at $T$ = 30\,K,
has been observed in magnetic susceptibility and heat capacity measurements
of \nco\/ however, there is no prior report of a concurrent structural
distortion. \cite{ishibashi_2007,Klemme_2002} In the current study, using
high-resolution X-ray powder diffraction, we find evidence for a structural
distortion at $T$ = 30\,K, as described in detail in a later section.
Similarly, an orthorhombic distortion of the already Jahn-Teller
distorted tetragonal \cco\/ occurs concurrently with ferrimagnetic ordering
at 130\,K (Fig.\,\ref{fig:vtcco}). This transition in \cco, not previously
noted in structural or diffraction studies, is observed here using
variable-temperature high-resolution synchrotron X-ray powder diffraction
performed on a sample of crushed single-crystals.  

\subsection{Crystal structure}

\begin{table*}
\caption{\label{tab:rietveld} 
Structural parameters of \nco\/ and \cco\/ obtained from Rietveld refinement
of high-resolution synchrotron X-ray diffraction data collected at
temperatures above and below the orthorhombic distortion of both compounds.}
\centering
\begin{tabular}{llllllll }
\hline
&& \nco\/ & & & & \cco\/ &  \ \\
\hline
&&Orthorhombic & Tetragonal & & & Orthorhombic & Tetragonal \	\\
\hline
\hline
$T$                        &  & 10\,K & 100\,K & & & 10\,K & 298\,K\ \\
Space group                &  & $Fddd$& $I4_1/amd$& & & $Fddd$ & $I4_1/amd$\ \\
Setting                    &  & origin 2 & origin 2 & & & origin 2 & origin 2\ \\
$Z$                        &  & 8 & 4 & & & 8 & 4\ \\
$a$ (\AA)                  &  & 8.18139(5) & 5.79029(2) & & & 7.71271(2) & 6.03277(1) \\
$b$ (\AA)                  &  & 8.16699(4) & 5.79029(2) & & & 8.53611(2) & 6.03277(1) \\
$c$ (\AA)                  &  & 8.56786(4) & 8.54639(4) & & & 8.54357(2) & 7.78128(1) \\
Vol/$Z$ (\AA$^3$)          &  & 71.5601(6)  & 71.6346(4)  & & & 70.3098(3)  & 70.7986(2) \\
Ni/Cu                      &  & $8a$ (1/8,\,1/8\,1/8) & $4a$ (0,\,1/4,\,3/8)& & 
                           & $8a$ (1/8,\,1/8\,1/8)& $4a$ (0,\,1/4,\,3/8) \\
$U_{iso}$ ($10^2$ \AA$^2$) & & 0.01(1) & 0.13(1)& & & 0.08(1)& 0.67(1) \\
Cr                         & & $16d$ (1/2,\,1/2, \,1/2) & $8d$ (0,0,0) & & & 
                               $16d$ (1/2,\,1/2, \,1/2) & $8d$ (0,0,0)  \\
$U_{iso}$ ($10^2$ \AA$^2$) & & 0.01(1) & 0.019(1)& & & 0.07(1) & 0.29(1) \\
O                          & & $32h$ ($x,y,z$) & $16h$ ($0,y,z$)& & & $32h$ ($x,y,z$) & $16h$ ($0,y,z$)\ \\
                           & & $x$ 0.2561(2)& $x$ 0& & & 0.2446(1) & 0 \\
                           & & $y$ 0.2589(2) & $y$ 0.5152(2)& & & 0.2675(2) & 0.5364(1)\ \\
                           &  &$z$ 0.2683(1) & $z$ 0.2322(2)& & & 0.2675(2) & 0.2526(1)\ \\
$U_{iso}$ (10$^2$ \AA$^2$) & & 0.03(2) & 0.16(2)& & & 0.06(2)& 0.55(1) \\
$\chi^2$                   & & 3.85 & 4.15 & & & 2.31 & 3.84 \\
$R_p$    (\%)              & & 6.25 & 7.06 & & & 7.50 & 8.96 \\
$R_{wp}$ (\%)              & & 8.39 & 9.41 & & & 8.39 & 6.65 \\
\hline
\hline
\end{tabular}
\end{table*}

The ambient temperature structure of both compounds can be indexed in the
tetragonal centrosymmetric space group $I4/amd$. At 298\,K, \nco\/ is still
undergoing the Jahn-Teller driven cubic-tetragonal transition and better
structural parameters of the tetragonal phase are obtained at 100\,K.
Structural parameters obtained from Rietveld refinement of 100\,K diffraction
data for \nco and 298\,K diffraction data for \cco\/ to the space group
$I4/amd$ are shown in Table\,\ref{tab:rietveld} and are in good agreement with
previous reports.\cite{crottaz_1997,Ueno_1999}

\begin{figure}
\centering\includegraphics[width=9cm]{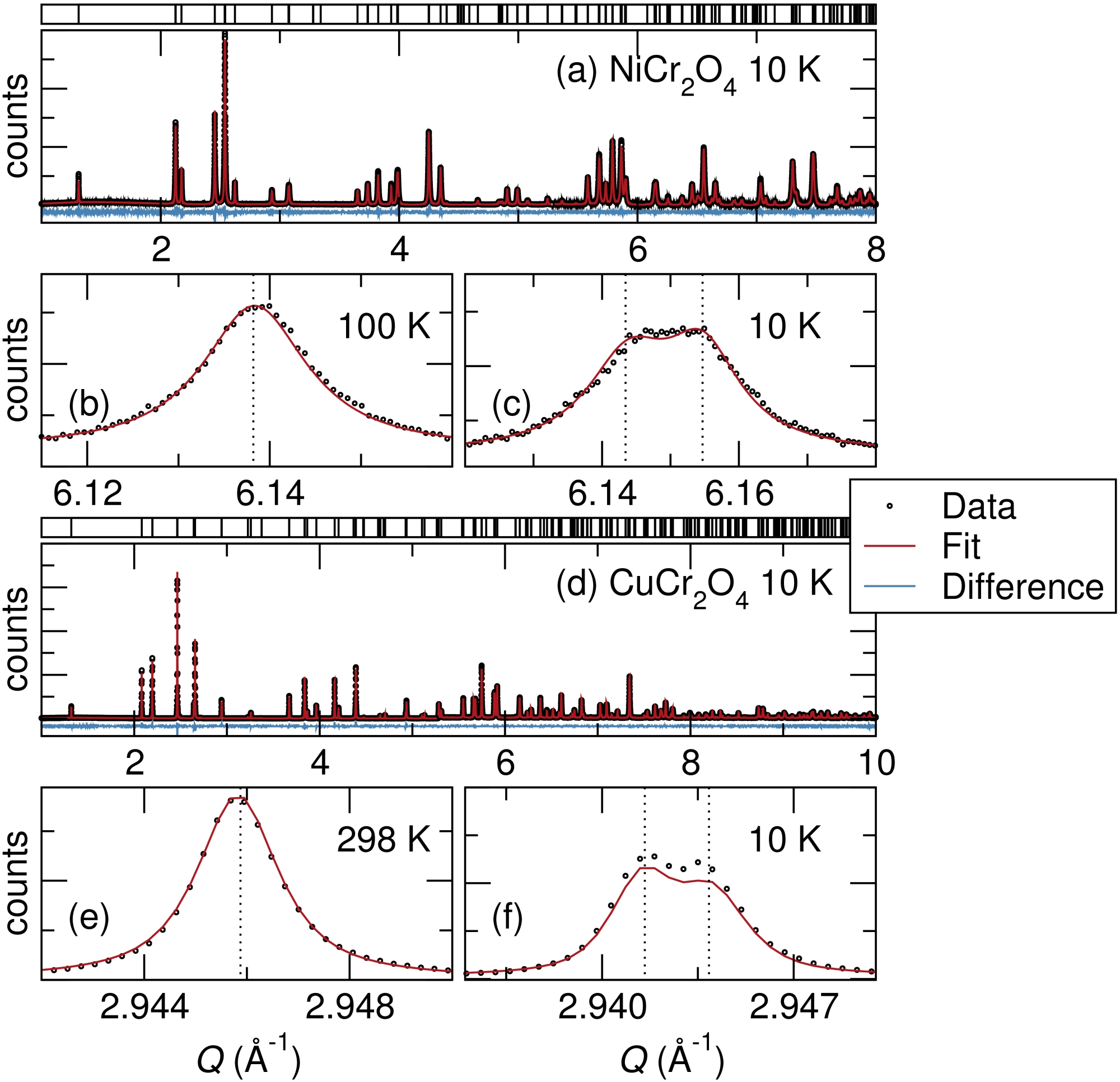}\\
\caption{(Color online) High resolution synchrotron powder X-ray diffraction
of \nco\/ and \cco. (a) The low temperature diffraction pattern of \nco\/ is
indexed to the orthorhombic space group $Fddd$. The lowering of average
crystal symmetry in \nco\/ from tetragonal to orthorhombic symmetry is
illustrated by the splitting of the (b) tetragonal (440) reflection into
(c) orthorhombic 800 and 080 reflections. (d) Like \nco, the
low temperature diffraction data of \cco\/ is indexed to the orthorhombic
space group $Fddd$ which is evident in the splitting of (e) (220) tetragonal
reflections into (f) 004 and 040 orthorhombic reflections. Structural
models are fit to the X-ray powder diffraction patterns using the Rietveld
refinement method.} 
\label{fig:structure}
\end{figure}

Magnetic ordering drives further structural distortions in \nco\/ and
\cco.\cite{bordacs_2009,ishibashi_2007} The low symmetry structures of
\nco\/ and \cco\/ are described by the orthorhombic spacegroup $Fddd$. $Fddd$
is a maximal nonisomorphic subgroup of $I4_1/amd$ and is derived from the
parent $Fd\bar{3}m$ by loss of all threefold rotation axes and part of the
twofold screw axes. Rietveld refinement fits of 10\,K diffraction data to the
orthorhombic space group $Fddd$ for both \nco\/  and \cco\/ are shown in
Fig.\,\ref{fig:structure}.  Symmetry lowering in \nco\/ and \cco\/ is
demonstrated by the splitting of certain high symmetry diffraction peaks as
illustrated in Fig.\,\ref{fig:structure} (c) and (f).  The current work is
the first description of the orthorhombic $Fddd$ structure for \cco. In
\nco, variable-temperature synchrotron X-ray diffraction measurements
show additional structural changes below 30\,K, in concurrence with anomalies
in specific heat and susceptibility measurements of \nco\/ reported both here
and previously in the literature.  This low temperature structural change of
\nco\/ is discussed in detail in a later subsection.

\begin{figure}
\centering\includegraphics[width=8cm]{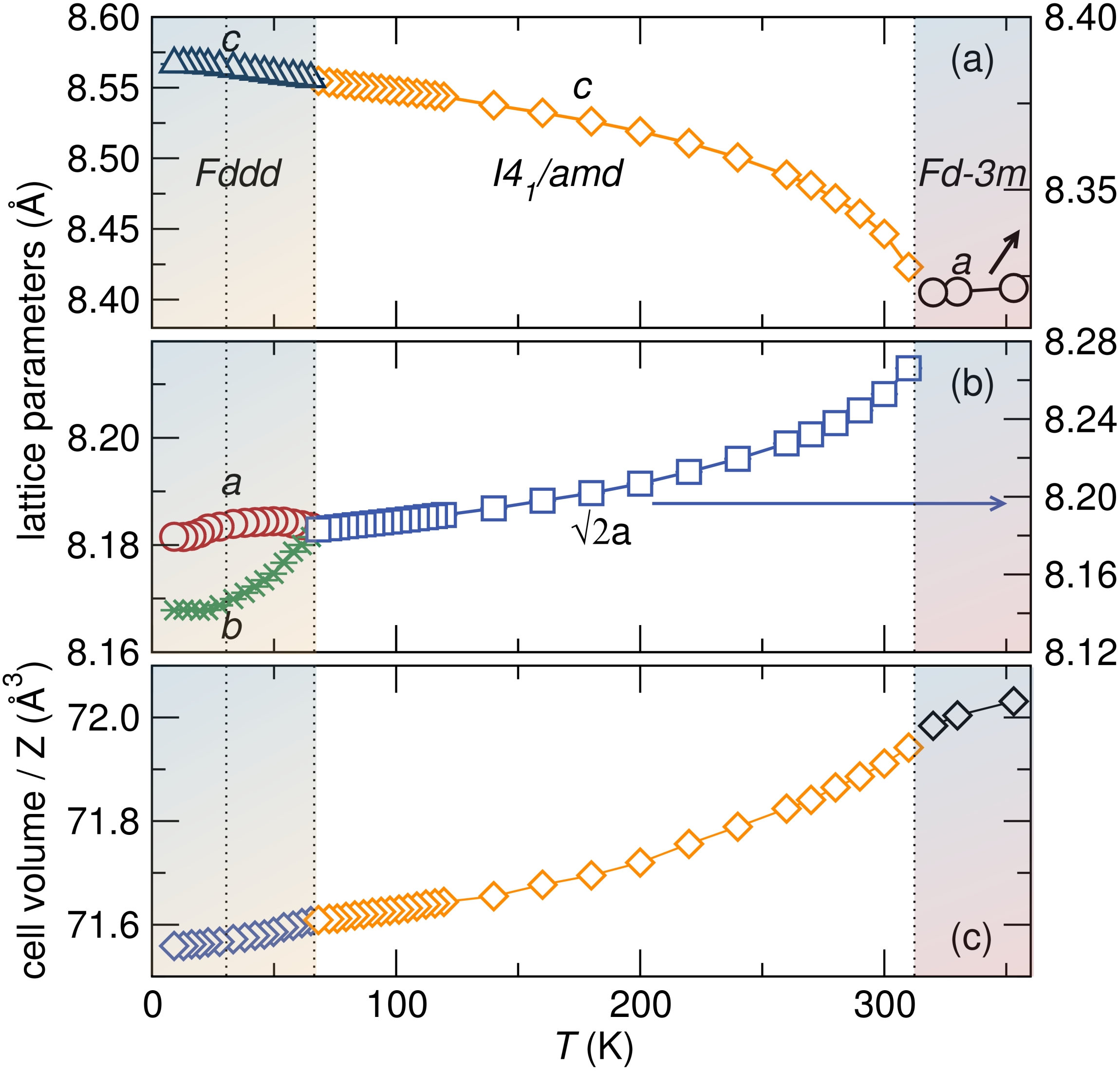}\\
\caption{(Color online) Changes in lattice parameters as a function of
temperature in \nco. (a) A cubic to tetragonal structural transition occurs
at 310\,K where the $a$ lattice constant of the cubic phase diverges into $a$
and $c$ lattice parameters of the tetragonal phase. The $a$ lattice constant
of the tetragonal cell is multiplied by $\sqrt{2}$ to clearly follow trends
in the lattice parameters of \nco. In the tetragonal phase, the $a$
parameter decreases (b) while $c$ increases (a) with decreasing temperature.
At 65\,K, a tetragonal to orthorhombic structural distortion occurs resulting
in three distinct lattice constants as shown in (a) and (b). (c) Variation of
the cell volume normalized by the number of formula units($Z$) in each cell.
A further structural distortion of orthorhombic \nco\/ occurs at 30\,K where
there is a slight discontinuity of the lattice parameters (a) and (b) and
cell volume (c); this is highlighted by the dashed line at $T$ = 30\,K.
In (a), (b) and (c) the error bars are smaller than the data symbols.}  
\label{fig:latticenco}
\end{figure}

\begin{figure}
\centering\includegraphics[width=8cm]{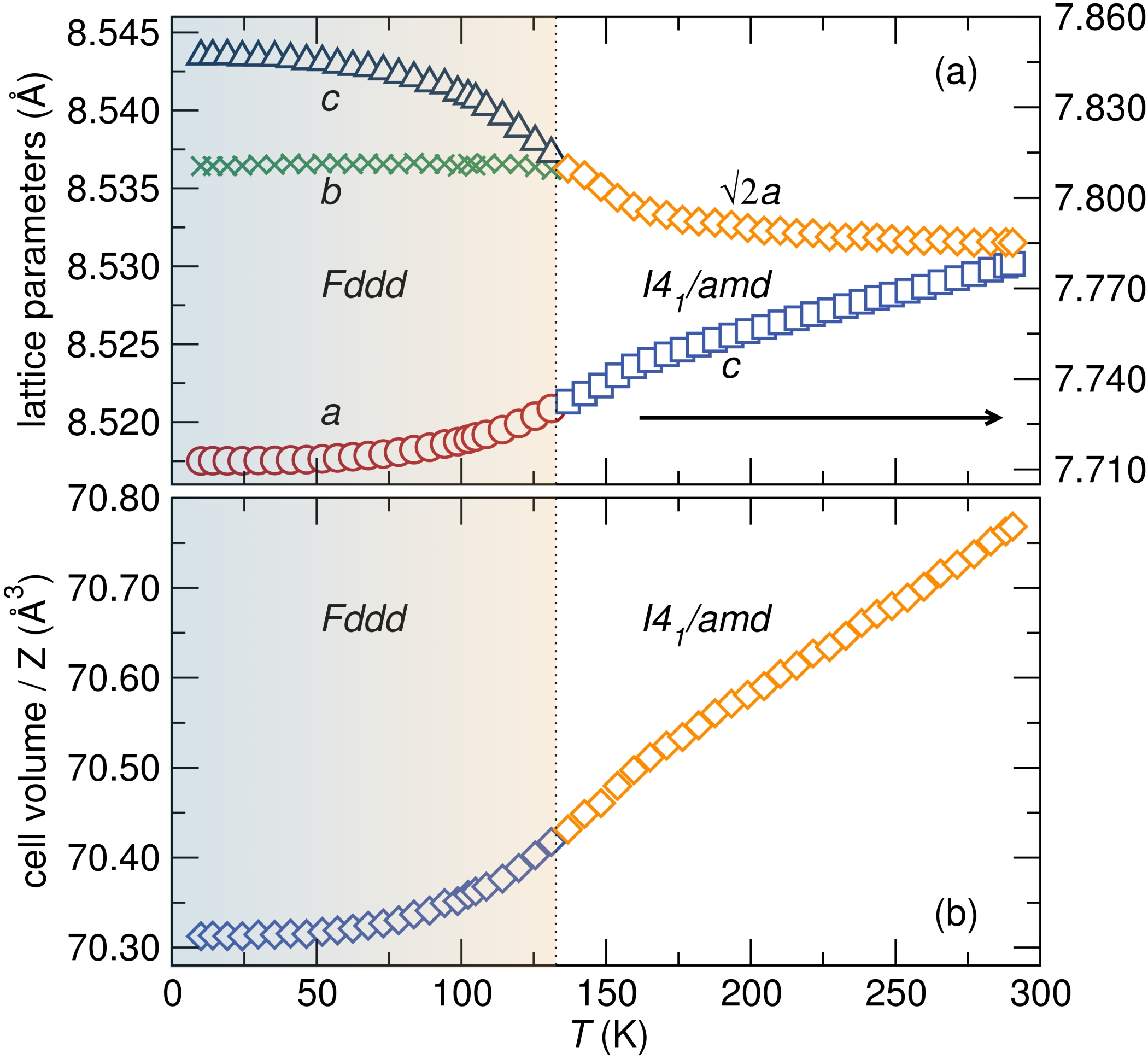}\\
\caption{(Color online) (a) The thermal evolution of lattice parameters of
\cco\/ reveals a tetragonal $I4_1/amd$ to orthorhombic $Fddd$ structural
transition at $\sim$ 130\,K. The tetragonal $a$ lattice parameter has been
multiplied by $\sqrt{2}$ to match the low temperature $b$ and $c$ lattice
values of the orthorhombic $Fddd$ cell. (b) Temperature dependence of the
cell volume normalized by the number of formula units ($Z$) in each cell
shows a steady decrease with temperature. In both (a) and (b), the error bars
are smaller than the data symbols.}  
\label{fig:latticecco}
\end{figure}

Changes in structural symmetry are reflected in the temperature dependence of
lattice parameters. At 310\,K there is a cubic to tetragonal transition in \nco\/ that
splits the cubic $a$ lattice constant into tetragonal $a$ and $c$ lattice
parameters [Fig.\,\ref{fig:latticenco} (a) and (b)]. Below 310\,K, the
tetragonal \nco\/ distortion grows, with an increasing $c$ and a decreasing
$a$ lattice constant (plotted as $\sqrt{2}a$). Below 65\,K, magnetic ordering
occurs concurrently with a transition to orthorhombic symmetry. The
tetragonal $a$ lattice parameter of \nco\/ diverges into distinct orthorhombic
$a$ and $b$ lattice constants [Fig.\,\ref{fig:latticenco}(b)]. At 30\,K, a
slope change clearly visible in the $a$ and $c$ lattice parameters 
[Fig.\,\ref{fig:latticenco}] matches anomalies in other property measurements 
as will be discussed later. \cco\/ is already tetragonal at ambient temperature 
due to cooperative Jahn-Teller ordering at 853\,K. The tetragonal lattice
constants of \cco\/ diverge below 300\,K with $c$ decreasing and the $a$
lattice constant [plotted as $\sqrt{2}a$ in Fig.\,\ref{fig:latticecco}(a)]
increasing, resulting in an enhanced tetragonal distortion with decreasing
temperature. Below 130\,K, where an orthorhombic distortion occurs
concurrently with the onset of ferrimagnetic order 
[Fig.\,\ref{fig:latticecco}(a)], distinct $a$, $b$, and $c$ orthorhombic 
lattice constants emerge. The orthorhombic lattice constants continue to 
diverge from 130\,K to the lowest temperatures measured as indicated in 
Fig.\,\ref{fig:latticecco}(a). The structural change due to orbital ordering 
in \nco\/ at 310\,K results in a discontinuity of the normalized cell volume
indicating a first order phase transition. In contrast, in the low
temperature tetragonal to orthorhombic phase transitions in \nco\/ and \cco\/
the continuous slope of the normalized cell volume through the
magnetostructural transition indicates a second order phase transition
[Fig.\,\ref{fig:latticenco}(c) and \ref{fig:latticecco} (b)]. 

\begin{figure}
\centering\includegraphics[width=8cm]{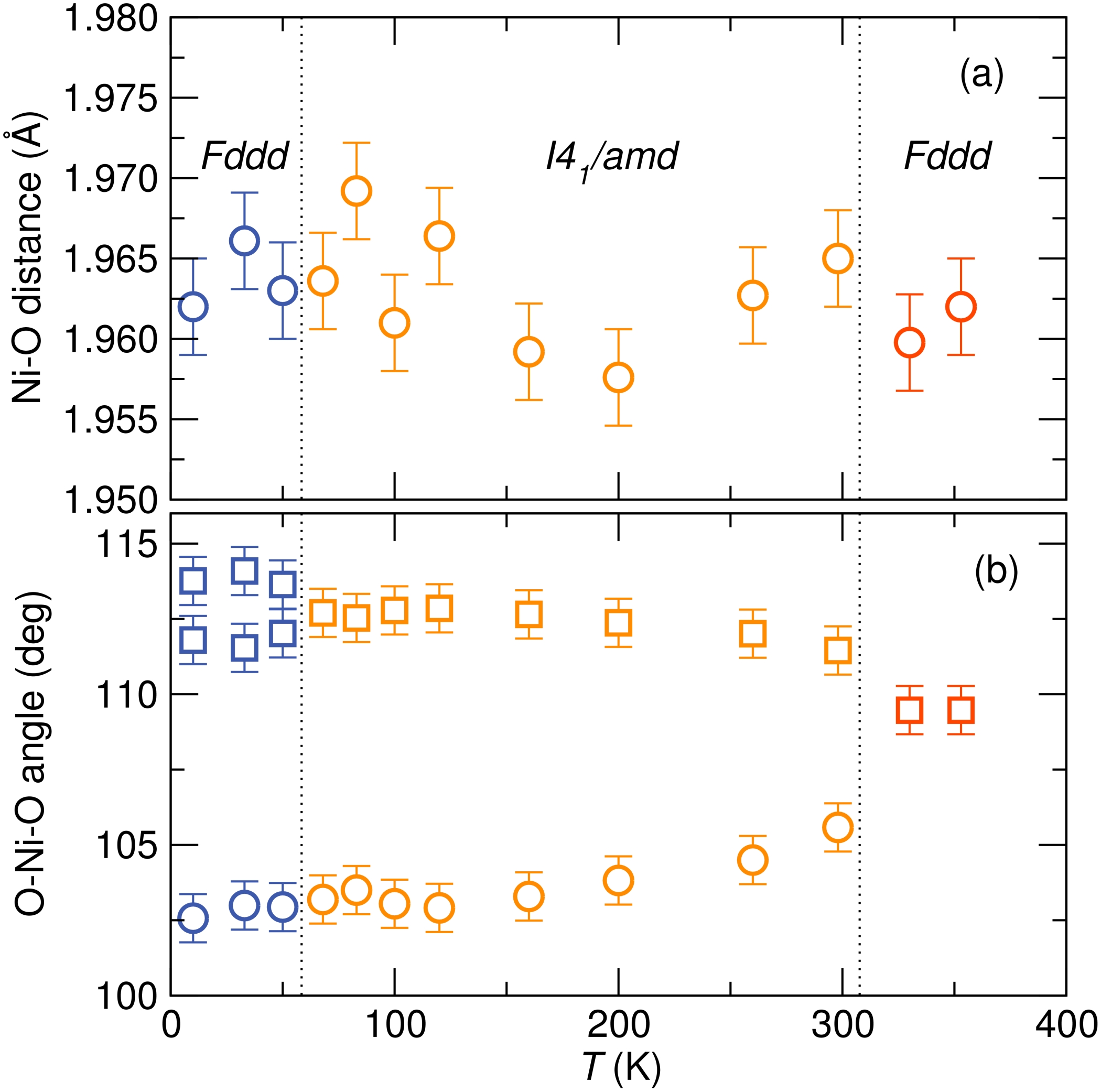}\\
\caption{(Color online) The variation in NiO$_4$ polyhedra as a
function of temperature. (a) The Ni-O bond length remains relatively constant
in all the structural phases (b) The single O-Ni-O angle of the cubic phase
separates into a larger angle and a smaller angle in the tetragonal phase.
Below the orthorhombic transition, there are three distinct O-Ni-O angles.}
\label{fig:NiO}
\end{figure}

\begin{figure}
\centering\includegraphics[width=8cm]{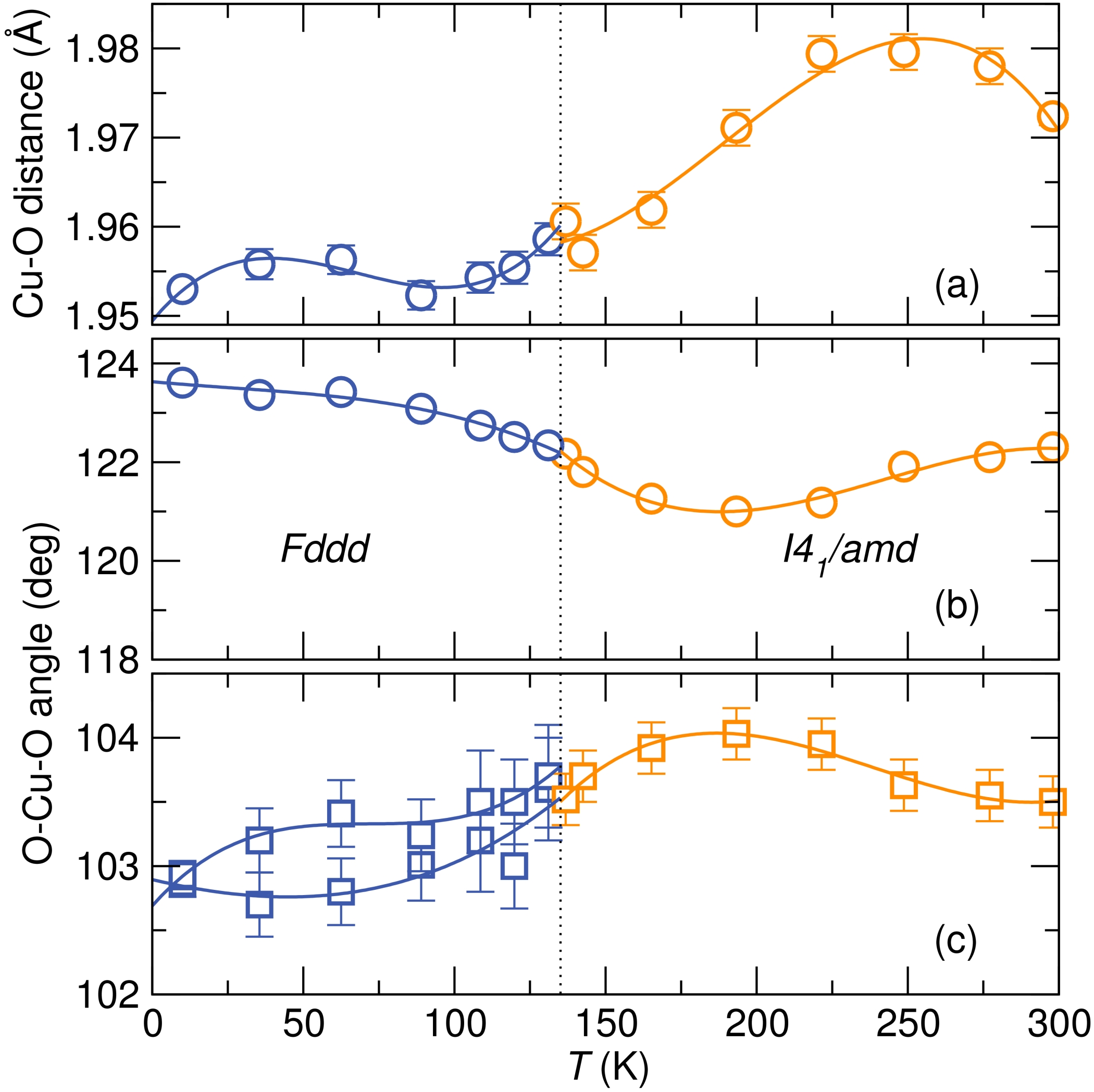}\\
\caption{(color online) Changes in the CuO$_4$ polyhedra as a function of
temperature (a) There is an overall decrease in the Cu-O bond distance, (b)
an increase in the larger O-Cu-O angle, and (c) a decrease in the smaller
O-Cu-O angle coupled with a splitting of this angle. These trends are
obtained from Rietveld refinement of synchrotron X-ray diffraction data.} 
\label{fig:CuO}
\end{figure}

\begin{figure}
\centering\includegraphics[width=9cm]{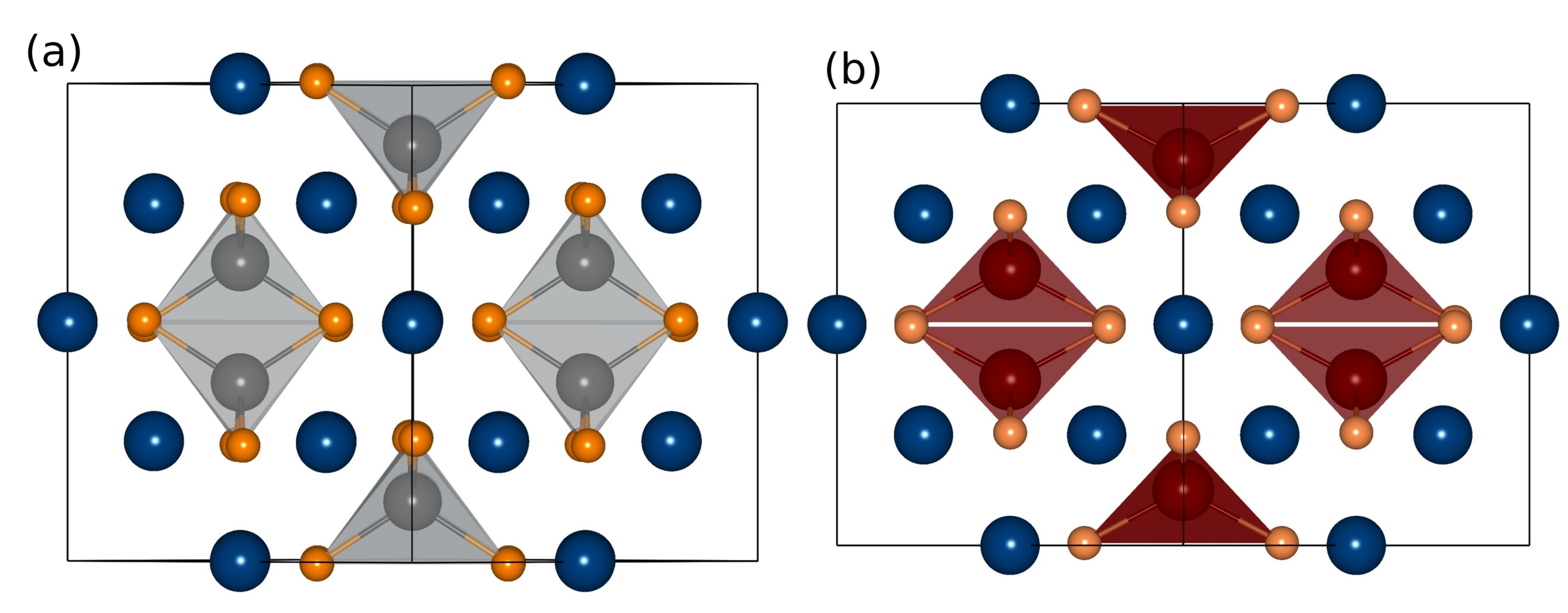}\\
\caption{(Color online) Low temperature (10\,K) orthorhombic crystal
structures of (a)\nco\/ and (b) \cco\/ projected down the [101] direction. 
Ni(grey) and Cu(red) are tetrahedrally coordinated by oxygen (orange).
Chromium is shown in blue. The elongation of NiO$_4$ tetrahedra along with
the compression of CuO$_4$ polyhedra is clearly seen in the low temperature
average structures.} 
\label{fig:vestastructure}
\end{figure}

Structural changes in \nco\/ and \cco\/ originate from deformations of
NiO$_4$ and CuO$_4$ polyhedra. In a perfect tetrahedron, all bond lengths are
equal and all O-Cation-O angles are 109.5$^{\circ}$. Ideal NiO$_4$ tetrahedra
are observed in the cubic \nco\/ structure above 310\,K [Fig.\,\ref{fig:NiO}
(a) and (b)]. Orbital ordering results in a distorted tetrahedron with a
single Ni-O bond distance, but two O-Ni-O angles [Fig.\,\ref{fig:NiO} (a)
and (b)] in the tetragonal phase. Below 65\,K, the orthorhombic structure
preserves a single Ni-O bond length, but splits the O-Ni-O angles into three
distinct O-Ni-O angles in the NiO$_4$ tetrahedra [Fig.\,\ref{fig:NiO} (a) and
(b)]. These distortions in Ni-O bond lengths and O-Ni-O bond angles result in
an elongation of NiO$_4$ tetrahedra.  At ambient temperature, CuO$_4$
tetrahedra are already significantly distorted with two different O-Cu-O
angles and a single Cu-O bond distance. With decrease in temperature and the
onset of the orthorhombic structural transition, we note a decrease in Cu-O
bond lengths [Fig.\,\ref{fig:CuO}(a)], an increase in the larger O-Cu-O
angle [Fig.\,\ref{fig:CuO}(b)] and a decrease in the smaller O-Cu-O angle
[Fig.\,\ref{fig:CuO}(c)]. The two smaller O-Cu-O angles divide into two. The
overall effect of these structural changes is a flattening of the CuO$_4$
polyhedra toward a square planar configuration. The differences in the
distortion of the CuO$_4$ and NiO$_4$ tetrahedra are apparent in the average
low temperature structures of \nco\/ and \cco\/ shown in 
Fig.\,\ref{fig:vestastructure}.

\subsection{Heat capacity}

\begin{figure}
\centering\includegraphics[width=8cm]{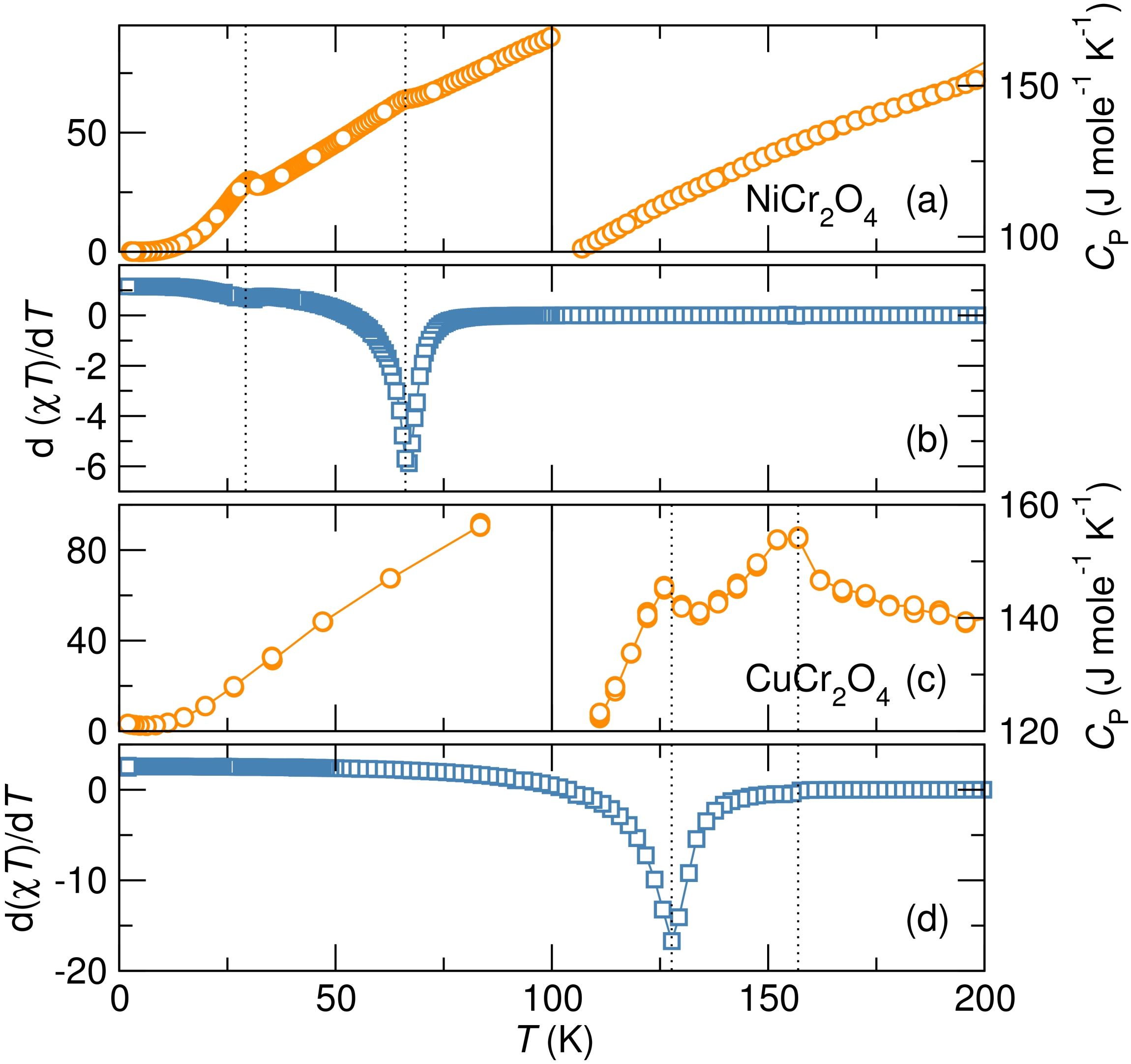}\\
\caption{(color online) (a) Entropy changes in \nco\/ and \cco\/ resulting
from structural and magnetic transformations.  (a) The heat capacity of
\nco\/ shows two anomalies at 65\,K and 30\,K. (b) Fisher heat capacity of
\nco\/ indicating release of magnetic entropy occurring at the same
temperatures where changes in specific heat are observed. (c) \cco\/ also
shows two transitions in the heat capacity at 155\,K and 130\,K. Concurrent
with these changes in heat capacity of \cco\/ are variations in magnetic
structure as illustrated by Fisher heat capacity shown in (d).}
\label{fig:hc}
\end{figure}

There are several interesting features in the heat capacity of \nco\/ and
\cco\/ that occur concurrently with magnetic and structural transformations
in these compounds.  Klemme and Miltenburg report three anomalies in the heat
capacity of \nco\/ occurring at 310\,K, 75\,K, and 30\,K.\cite{Klemme_2002}
Our heat capacity measurements over the temperature range 3\,K $ \leq T \leq$
200\,K for \nco\/ show two anomalies at 65\,K and 30\,K 
[Fig.\,\ref{fig:hc}(a)]. 
The Jahn-Teller cubic-tetragonal structural distortion of
\nco\/ causes the anomaly in heat capacity at 310\,K reported by Klemme and
Miltenburg.\cite{Klemme_2002} The transition into a ferrimagnetic ordered
state [Fig.\,\ref{fig:vtnco}(a)] that occurs concurrently with a structural
change [Fig.\,\ref{fig:vtnco}(b)] results in the change in entropy that we
observe at 65\,K and was reported by Klemme and Miltenburg to occur at
$T$ = 75\,K. Klemme and Miltenburg also reported an additional anomaly in
specific heat at 30\,K; Ishibashi and Yasumi noted a change in magnetic
susceptibility at this temperature. \cite{Klemme_2002,ishibashi_2007} We
observe this anomaly in the heat capacity of \nco\/ at 30\,K and attribute it
to an additional change in the magnetic [Fig.\,\ref{fig:hc}(b)] and crystal
structure [Fig.\,\ref{fig:vtnco}(b)] as will be discussed in section D of
this paper.

There are two anomalies in the specific heat of \cco\/ at 130\,K and 155\,K
[Fig.\,\ref{fig:hc}(c)]. The anomaly at 130\,K is coincident with
ferrimagnetic [Fig.\,\ref{fig:hc}(d)] and tetragonal-orthorhombic 
[Fig.\,\ref{fig:vtcco}(b)] phase transitions in the compound. The 
transition into the orthorhombic ferrimagnetic state in \cco\/ occurs through 
an intermediate step with signatures in Fisher heat capacity and specific heat 
measurements at 155\,K 
[Fig.\,\ref{fig:hc} (c) and (d)].\cite{fisher_1962} Slight
structural effects accompany this second transition as shown in 
Fig.\,\ref{fig:latticecco}(b) where there is a subtle inflection point of the
evolution of cell volume with temperature. Further characterization of this
intermediate change in the magnetism of \cco\/ at about 155\,K requires
careful investigation in future study.

\subsection{30\,K magnetostructural transition of NiCr$_2$O$_4$}

\begin{figure}
\centering\includegraphics[width=8cm]{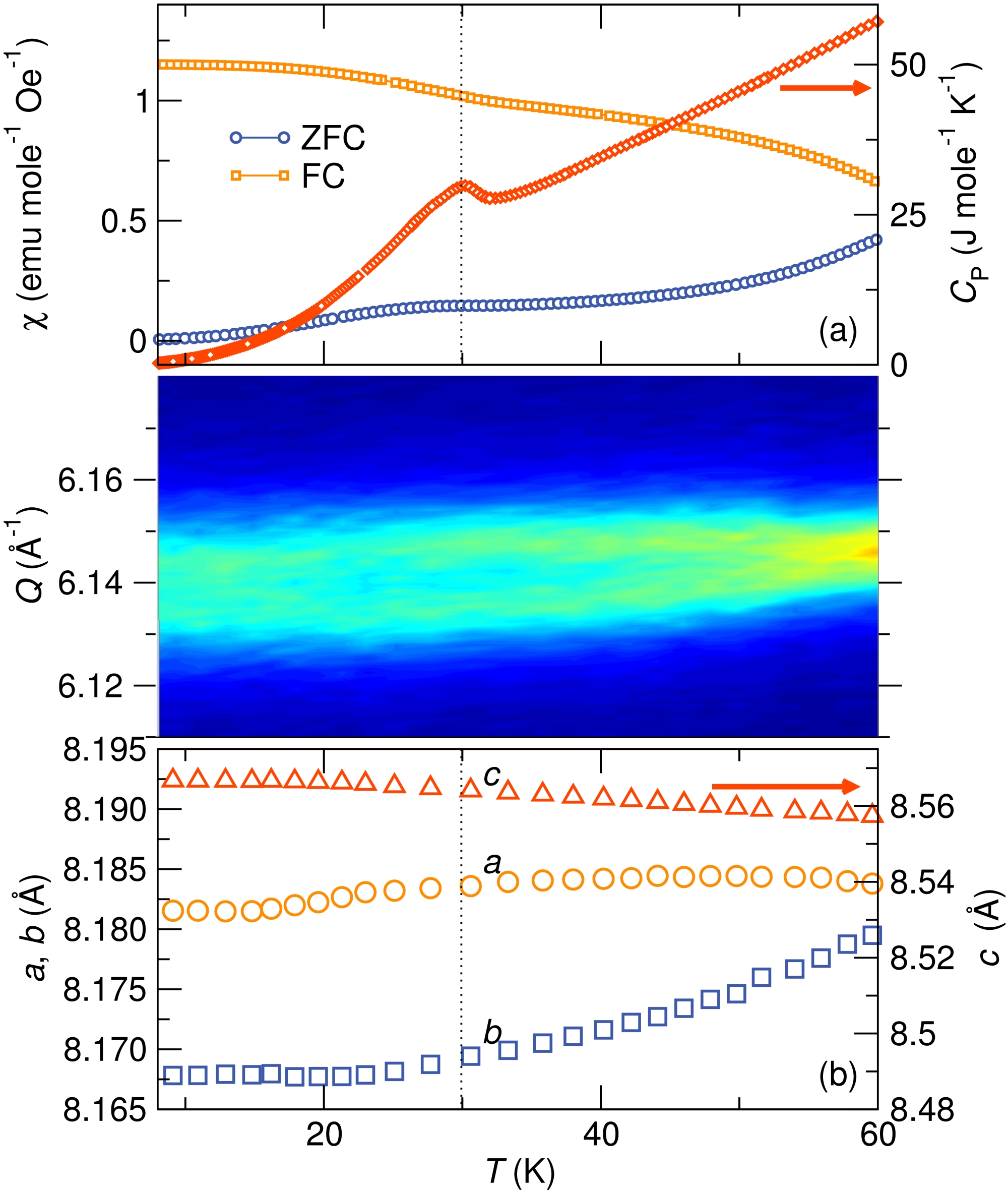}\\
\caption{(color online) Changes in magnetic order and heat capacity of \nco\/
at $T$ = 30\,K are accompanied by a structural change. (a) Zero field
cooled and field cooled temperature dependent magnetic susceptibility
measurements of \nco\/ show a change in magnetic order at $T$ = 30\,K.
Concurrent with this transition in magnetism is a change in entropy indicated
by the anomaly in heat capacity. The central panel tracks changes in intensity
of the orthorhombic 080 and 800 reflections  at $T$ = 30\,K
illustrating that a structural change takes place at $T$ = 30\,K. (c) This
structural change is also reflected in the temperature dependent lattice
constants of \nco\/ which vary at this temperature.} 
\label{fig:NCOlt}
\end{figure}

During the ferrimagnetic transition of \nco, a simultaneous cooperative
crystal distortion from tetragonal to orthorhombic symmetry occurs as
reported by Ishibashi and Yasumi.\cite{tomiyasu_2004, ishibashi_2007} We
observe this magnetostructural transition in \nco\/ at $T$ = 65\,K 
(Fig.\,\ref{fig:vtnco},\ref{fig:latticenco}). 
Magnetic susceptibility measurements
by Ishibashi and Yasumi show yet another low temperature magnetic transition
in \nco\/ at $T$ = 31\,K that was reported by Tomitasu and Kagomiya as
corresponding to the ordering of the antiferromagnetic component of the
magnetic structure of \nco. Klemme and Miltenburg also observed a change in
entropy at this temperature\cite{Klemme_2002}, however, no changes of the
average structure of \nco\/ have been observed at $T$ = 31\,K.
\cite{tomiyasu_2004, ishibashi_2007} Our measurements reveal similar
anomalies in the magnetism and specific heat measurements of \nco\/
[Fig.\,\ref{fig:NCOlt}(a)] at $T$ = 30\,K. Furthermore, we observe a slight
change in average structure at this temperature. The central panel in 
Fig.\,\ref{fig:NCOlt}(b) tracks a \nco\/ Bragg diffraction peak as a function 
of temperature and shows a distinct peak narrowing and intensity change below
30\,K. Likewise, the $Fddd$ lattice parameters obtained from Rietveld
analysis of the variable temperature diffraction data 
Fig.\,\ref{fig:latticenco} show a clear change in slope near 30\,K 
[Fig.\,\ref{fig:NCOlt}(b)]. However, no evidence for a further change of \nco\/
symmetry (\textit{eg.} to monoclinic) below 30\,K is found in these 
high-resolution powder diffraction data. This is the first report of a 
structural effect concurrent with reported anomalies in heat capacity and 
magnetic measurements, and will be further examined in future studies.

\section{Conclusions}

Structural changes occur concurrently with magnetic phase transitions in
\nco\/ and \cco. We have resolved details of the crystal structure of 
the low temperature phase of \nco\/ and
\cco\/ in the orthorhombic space group $Fddd$ and present the first
structural description of orthorhombic \cco. We find that the magnetic
transition at 30\,K in \nco\/ is also accompanied by further, subtle structural
anomaly. Pronounced elongation of NiO$_4$ tetrahedra, and compression of
CuO$_4$ tetrahedra toward a square planar configuration drive the distortions
into the orthorhombic phase in these compounds. As postulated by
Smart and Greenwald, we suggest that multiple exchange coupling pathways in
the distorted orthorhombic structure are likely to be the reason behind the 
strong magnetostructurtal coupling observed in these compounds.\cite{smart1950} 
We anticipate that this study will inspire further investigation of such 
coupling in ferrimagnetic spinels.

\section{Acknowledgements}

The 11-BM beamline at the Advanced Photon Source is supported by the 
Department of Energy, Office of Science, Office of Basic Energy Sciences, 
under contract no. DE-AC02-06CH11357. 
MCK thanks P. T. Barton and A. Goldman for helpful discussions. 
MCK is supported by the Schlumberger Foundation Faculty for the Future 
Fellowship, and the research (MCK and RS) is supported by the National Science 
Foundation through a Materials World Network grant (DMR 0909180). 
We acknowledge the use of  shared experimental facilities of the Materials 
Research Laboratory: an NSF MRSEC, supported by NSF DMR 112105. The MRL is a 
member of the NSF-supported Materials Research Facilities Network 
(www.mrfn.org). 

\bibliography{Kemei}

\end{document}